# *FeCaffe*: FPGA-enabled Caffe with OpenCL for Deep Learning Training and Inference on Intel Stratix 10

Ke He, Bo Liu^, Yu Zhang, Andrew Ling* and Dian Gu

IoTG Vision Market Channel PRC, ^Flex Services and *Programmable Solution Group of Intel Corporation

{harvey.he | bo.a.liu | richard.yu.zhang | andrew.ling | penny.gu}@intel.com

## ABSTRACT

Deep learning and Convolutional Neural Network (CNN) have becoming increasingly more popular and important in both academic and industrial areas in recent years cause they are able to provide better accuracy and result in classification, detection and recognition areas, compared to traditional approaches. Currently, there are many popular frameworks in the market for deep learning development, such as Caffe, TensorFlow, Pytorch, and most of frameworks natively support CPU and consider GPU as the mainline accelerator by default. FPGA device, viewed as a potential heterogeneous platform, still cannot provide a comprehensive support for CNN development in popular frameworks, in particular to the training phase. In this paper, we firstly propose the *FeCaffe*, i.e. FPGA-enabled Caffe, a hierarchical software and hardware design methodology based on the Caffe to enable FPGA to support mainline deep learning development features, e.g. training and inference with Caffe. Furthermore, we provide some benchmarks with FeCaffe by taking some classical CNN networks as examples, and further analysis of kernel execution time in details accordingly. Finally, some optimization directions including FPGA kernel design, system pipeline, network architecture, user case application and heterogeneous platform levels, have been proposed gradually to improve FeCaffe performance and efficiency. The result demonstrates the proposed FeCaffe is capable of supporting almost full features during CNN network training and inference respectively with high degree of design flexibility, expansibility and reusability for deep learning development. Compared to prior studies, our architecture can support more network and training settings, and current configuration can achieve 6.4x and 8.4x average execution time improvement for forward and backward respectively for LeNet.

## KEYWORDS
FeCaffe, FPGA, Deep Learning, CNN, Training, Inference, Caffe, OpenCL, Heterogeneous Platform, Stratix 10

## 1 Introduction

Deep learning has becoming increasingly more popular and drawn huge attention in both academic and industrial areas in recent years. The Convolutional Neural Network (CNN), as the subset of deep learning, has already demonstrated the capability for higher accuracy in classification, detection and recognition areas, compared to traditional computer vision methods, and thus it has been widely applied to commercial markets, e.g. digital secure surveillance, retail, industrial areas etc.

With rapid development of deep learning and CNN technology, the framework also has gained sufficient attention and investments to improve and develop as well. The development of CNN network is a sophisticated and systematic process, and it usually contains dataset preparation, pre/post processing, training, validation, and acceleration with heterogeneous platforms etc. All of these actions are required by using the framework so that deep learning algorithm developers can focus on algorithm development only with ease.

Currently, most of popular deep learning frameworks natively support Central Processing Unit (CPU) and consider the Graphic Processing Unit (GPU) as the accelerator by default. Field Programmable Gate Array (FPGA) viewed as another potential device by nature for heterogeneous platform acceleration, the development approach is still comparatively sophisticated, and thus cannot be comprehensively supported by popular frameworks for CNN development, especially in terms of training.

In this paper, in order to improve such a situation to some extent, we propose *FeCaffe*, i.e. FPGA-enabled Caffe framework with OpenCL and provide some contributions as follows:

- Seamlessly integrate FPGA into Caffe framework to perform CNN network training. To our best knowledge, it is the first time to enable FPGA to provide training features for popular networks and support entire training process and various training settings with Caffe.
- Introduce hierarchical software and hardware architectures in details, and the proposed approach has potential to expand to other OpenCL-backend frameworks due to OpenCL portability and generality.
- The proposed FeCaffe has high degree of design flexibility in terms of novel function development and integration, achieving same fine-grained level with GPU. Following the proposed approach, users can develop new kernels and integrate them into the FeCaffe conveniently. Moreover, various FPGA processing architectures can also be integrated to improve computation efficiency as required.
- Compared to prior work, the proposed approach is able to support more CNN network topologies, training-related settings, and provide better expansibility and ease of use[8] [9] . With regard to the performance, current configuration can achieve 6.4x and 8.4x average improvement for forward and backward respectively for LeNet under same testing conditions [8] .

- First time to support SqueezeNet and GoogLeNet training process with default or customized training settings on FPGA, in supporting multiple loss function definitions. In addition, we also firstly provide the benchmark of training an epoch based on ImageNet 2012 training and validation dataset for SqueezeNet and GoogLeNet respectively.

The rest of this paper is organized as follows. In Section 2, Caffe framework, FPGA OpenCL development, and deep learning with FPGA are introduced respectively. Section 3 describes the design methodology, including hierarchical software and hardware architecture, memory synchronization mechanism. Section 4 presents the result and Section 5 provides the analysis and optimization directions accordingly. Eventually, this paper is concluded in Section 6.

## 2 Related Work

### 2.1 Caffe Framework

Among deep learning frameworks in the market, Caffe, standing for Convolutional Architecture for Fast Feature Embedding, has been viewed as one of the most popular and important deep learning frameworks [1] . Original Caffe natively supports operations on CPU with a number of libraries, e.g. Basic Linear Algebra Subprograms (BLAS) and Math Kernel Library (MKL), and also NVidia GPU as the default accelerator with Compute Unified Device Architecture (CUDA) programming or CUDA Deep Neural Network (CuDNN) library. Some classical and well-known CNN networks, e.g. AlexNet, VGG, GoogLeNet, SqueezeNet, etc., were developed and further widely applied in many applications and scenarios by using Caffe[4] [5] [6] [7] .

### 2.2 OpenCL and FPGA Development

Register Transfer Level (RTL) coding, e.g. Verilog and VHSIC Hardware Description Language (VHDL), has been considered as the conventional FPGA development languages for a long history. It is a hardware-oriented and efficient approach, but requires massive engineering development efforts and comprehensive underlying details of FPGA circuit and design flow skills, e.g. synthesis, placement and routing to achieve a good result in terms of performance and timing. In addition, conventional FPGA development flow does not have a friendly simulation environment, especially for algorithm development. With the increment of size and complexity, in particular to the deep learning and CNN applications, those disadvantages of RTL designs are becoming increasingly more obvious. In order to address these pain points, FPGA vendors provide high-level language design methodology and tools for FPGA development, such as High-level Language Synthesis (HLS) and OpenCL [22] [22] [24] [25] .

OpenCL is public standard with data and task parallel programming models, initially proposed by *Khronos Group*, especially for parallel acceleration on heterogonous platforms, e.g. GPUs, CPUs and FPGAs. The OpenCL design flow has two design stages: kernel and host development. Host part development is mainly used for device initialization, setup, managing memory allocation, and coordinating kernel behaviors. In this work, we refer to the host code as *runtime* functions and further divide the runtime into two groups: kernel-related and common runtime. The purpose of *common runtime* is to create context, command queue, program, and memory allocation while the *kernel-related runtime* focuses on kernel argument configuration, debug, profiling, launch and release. Kernel development means to develop offload functions and tune performance on various devices, and has two approaches: *NDRange* and *single work-item*. NDRange is the default execution model for OpenCL kernel development, and employs a number of build-in functions to complete mapping of algorithm to massive work-items execution concurrently. Single work-item is another design philosophy that is hardware-oriented methodology, achieving maximum throughput and optimization by using more flexible optimization directives, FPGA native components and deeper processing pipeline. Optimizing the system and performance with NDRange design approach is hardware-agnostic, by tuning parameters of group size or compute units, and thus it is general and universal for various devices because the compilers can manage resource and adjustment for each device automatically. On the contrary, single work-item optimization heavily relies on complier tools and specific hardware architectures provided by various device vendors, and users' skillset as well.

### 2.3 CNN Inference with FPGA

Due to rapid growth in wide application areas, there have been a number of research studies based on FPGA for deep learning and CNN applications[11] [12] [13] [14] [15] [16] [17] [18] [19] [20] [21] . Authors in these papers have demonstrated FPGAs are able to achieve impressive benchmarks for some popular CNN networks on Intel and Xilinx devices with HLS, OpenCL and RTL design methodologies respectively. In general, they concentrate on the network inference efficiency, and thus defined their own processing architectures and pipelines with FPGA dedicated DSP blocks, distributed and BRAMs, to realize key CNN processing operations, e.g. convolution, pooling, in parallel with tiles simultaneously. Besides, some optimization technologies, for example, fixed-point quantization, low precision, and data transformation, e.g. Winograd and Fast Fourier Transform (FFT) have also been considered as the preconditions to realize such significantly competitive benchmarks on FPGAs, compared to GPUs and CPUs[14] [19] [20] [21] . Those low precision data type, e.g. Int8, is able to increase DSP efficiency, decrease the weight, intermediate and final result storage based on FPGA limited on-chip memory, and DDR bandwidth. The core of processing architectures is to utilize a number of cascaded DSP blocks to perform convolution or matrix multiplication in parallel in several dimensions. In general, the data movement path and processing architecture mechanisms are well-optimized and fixed down to achieve impressive results in terms of throughput, performance per power and energy efficiency, compared to CPUs and GPUs respectively. In addition, there are a few of studies on FPGA CNN inference with Caffe framework. Authors in [10] and [16] presented both hardware architectures and software approaches for Caffe invocation and provided benchmarks on FPGA inference with Caffe for AlexNet, VGG, GoogLeNet and YOLO-v2 respectively.

### 2.4 CNN Training with FPGA

Compared to the inference with FPGA, the research of FPGA training is relative limited, and there are only two studies providing the implementation details and benchmarks. Authors in [8] proposed the pipeline structure of convolution and pooling layers and benchmarked the training process with two FPGA boards on LeNet. A much more aggressive approach based on FPGA clusters for CNN training has been proposed in [9], and 15 FPGAs, scaling up to 83 FPGAs at most, are used for training AlexNet, VGG-16 and VGG-19 respectively. They both employ multiple FPGA boards for CNN training, and create dedicated processing pipeline with fixed weight update mechanism. In addition, they both utilize customized runtime with low-level network configuration parameters and hardware constraints as the software control during the training, resulting in further limitation on CNN training usage.

### 2.5 Motivation

Most of previous studies only focus on CNN inference with FPGA while the training process of deep learning on FPGA has gained little attention. Among the FPGA-enable inference designs, the trend is to design the most efficient and dedicated processing architectures with fine-tuned and well-designed data buffer and reuse mechanism, e.g. data sharing, weight sharing or even hybrid, for one or some types of classical network topologies. This kind of design philosophy leads to the maximum FPGA throughput and efficiency for inference, but have to suffer from flexibility and adaptation problems in some practical CNN scenarios. For the inference structure, it is usually difficult and time consuming to insert new developed functions or primitives into the well-optimized pipelines, resulting in slow time to market and many development efforts for FPGA-based CNN solutions. CNN training with FPGA is more challenging than inference in terms of not only hardware designs and utilizations, but also software development. Due to high development barrier and a large amount of engineering efforts, only a few of studies are able to provide FPGA approaches with Caffe for inference and thus they are not complete approaches because training parts are excluded. There are still some gaps providing more functions and flexibility with FPGA for deep learning development, compared to GPUs. In summary, FPGA-based architectures have obvious limitations and gaps in terms of flexibility, customization and convenience for CNN training and inference development

Considering all of factors discussed above, we propose the FeCaffe in this paper, and make the contributions as discussed previously. This study is a more comprehensive approach and constitutes an extension to conventional CNN development, which often considers GPUs and CPUs, and also creates more feasibilities and choices for deep learning development based on FPGA-related heterogeneous platforms.

## 3 Design Methodology

### 3.1 Caffe Architecture

Conventional Caffe framework structure is illustrated in Figure 1, we divide the whole hierarchy into six layers: from network application level to hardware device layer. Note that we only describe some hierarchical functions that are related to hardware devices and CNN operation layers because Caffe framework also has a larger number of components on debug and logging, database I/O processing, and protobuf parsing, etc. Those components can often be reused with almost no change for various Caffe variants.

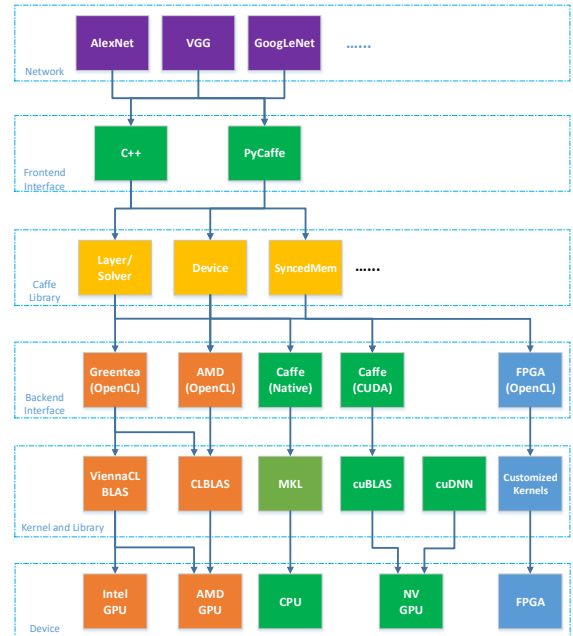

**Figure 1 Hierarchical Architecture of Caffe Framework**
(Green: native Caffe provided; Red: GPU-based OpenCL variants provided; Orange: Hardware-related functions; Blue: Our work)

Training and inference for various networks can be performed by calling for C++ and Python interface with native Caffe support. Either C++ or Python can call for the Caffe libraries consisting of a large number of defined *classes*, e.g. *layer*, *device*, *syncedmem* etc. For the CPU path, it can start from either C++ or Python interface and invoke some math functions or perform some operations directly defined by the layer functions with C++. Those math functions further calls for MKL or BLAS libraries, and finally maps to CPU device. Similarly, GPU approach goes CUDA interface and invokes the math functions optimized by cuBLAS or cuDNN, and some CUDA functions for layer operations. Some hardware-related classes, e.g. device and syncedmem, are mainly used for GPU device and memory management.

Due to open source of Caffe framework and community contributions, a number of OpenCL-based variants, stemmed from native Caffe, have been proposed and maintained, and thus they are capable of supporting more heterogeneous platforms, e.g. AMD and Intel integrated GPUs, among which, two well-known represetives are analyzed in this paper [2] [3]. Author in [2] proposed an OpenCL-based interface mechanism named *greentea*, and, provided the CNN acceleration library with OpenCL by leveraging official CLBLAS and Vienna CLBLAS libraries. With good compatibility of those libraries, the greentea is able to support CNN activities within Caffe on Intel integrated GPUs and AMD GPUs, i.e. the Greentea path in Figure 1. Similarly, another branch maintained by AMD proposed their backend interface hierarchy and kernel designs to support CNN operations deployed and optimized on AMD GPU devices, i.e. AMD path in Figure 1. It is important to note that some hardware-sensitive classes or functions, e.g. operation layers, device, syncedmem, highlighted by orange in Figure 1, also require significant modifications to support new devices even following the same OpenCL development flow.

### 3.2 Kernel-related Layers

Following the similar structure, we proposed a novel hierarchical backend interface based on OpenCL flow to support CNN operations on FPGA, i.e. labeled with blue in Figure 1. More details of our path for the kernel development and backend interface can be referred in Figure 2. Here we divide three layers from FPGA kernel development to deep learning operations within Caffe framework. The *L1*, i.e. *kernel layer*, includes all of kernel design files to support necessary operations and all of kernel files are compiled by Intel FPGA SDK for OpenCL and generate a FPGA configuration file. In order to support deep learning training and inference features, we group all of kernels required into three types: *layer-related*, *BLAS-related* and *solver-related*. The layer-related kernels define the functions to support some layer classes directly, e.g. pooling, activation functions, including both forward and backward operations. BLAS-related group contains some general and common functions from BLAS library, e.g. General Matrix Multiplication (GEMM), General Matrix-Vector Multiplication (GEMV), etc. Solver is employed to update the weights according to various approaches or policies during the training iterations, and thus plays a significant role during network training process. Some common weight update approaches, e.g. Stochastic Gradient Descent (SGD), *Adam*, *AdaDelta*, *Nesterov*, etc., are supported on FPGA. In this study, note that all of kernel files mentioned utilize NDRange design style from the open-sourced OpenCL Caffe versions and CLBLAS library for simplicity and saving validation energy.

For the upper layer on top of the kernel, this layer is referred to as the *L2 wrapper layer*, constituting a number of runtime functions corresponding each kernel from various groups in kernel layer. The runtime functions in this layer are all kernel-related runtimes as described previously, containing the kernel creation, argument setting, kernel launch, release and debug information. The main purpose of wrapper layer is further to encapsulate the OpenCL kernel designs for invocation with ease by high-level functions. With regard to the *L3*, *class layer*, important classes have been defined by native Caffe, but need to extend based on the underlying L1 and L2 layers accordingly. Some straightforward layers, e.g. pooling, can call for the corresponding runtime directly to realize acceleration on FPGA. Some functions and layers might need a combination of BLAS library and kernels, and thus leverage various runtime configurations accordingly. Following the proposed architecture and partitions, the proposed design methodology has the potential to be applied in other deep learning frameworks, e.g. TensorFlow or Pytorch. L1 and L2 structures can remain the same due to OpenCL common standard while only L3 is required to update according to high-level functions defined by various frameworks.

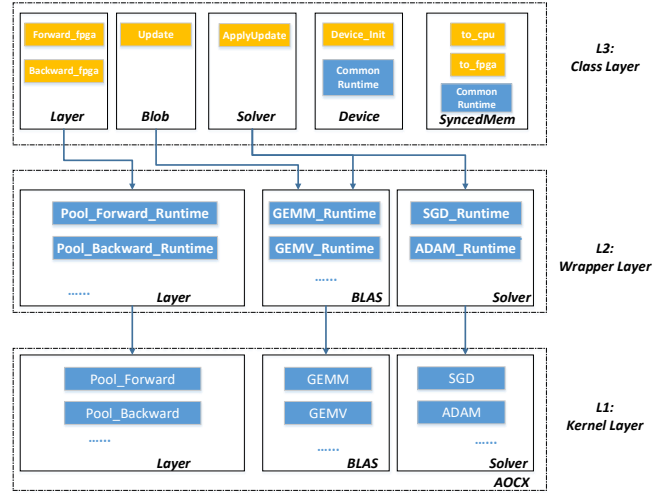

**Figure 2 Hierarchical FeCaffe Structure in Details**

### 3.3 Memory Synchronization and Fallback

Memory management is a great feature of Caffe framework, and is capable of allocating memory on-demand for efficient usage at both host and device side, and performing synchronization as required. Following this design idea, FeCaffe makes an extension to the scenario of memory management on FPGA. The memory status topography is shown in the top part of Figure 3.

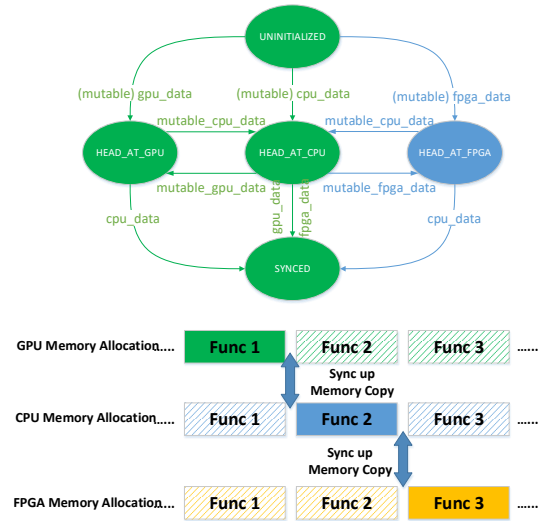

**Figure 3 Top: Memory Status Topography; Bottom: Workload Partition Configurations**

Table 1: Performance Benchmark with Native Caffe Time Measurement (Batch Size = 1)

| AlexNet (ms) | | | VGG_16 (ms) | | | SqueezeNet_v1.0 (ms) | | | GoogLeNet_v1 (ms) | | |
|---|---|---|---|---|---|---|---|---|---|---|---|
| Layer | Forward | Backward | Layer | Forward | Backward | Layer | Forward | Backward | Layer | Forward | Backward |
| Data | 0.001 | 0.001 | data | 0.002 | 0.002 | data | 0.001 | 0.001 | data | 0.635 | 0.003 |
| conv1 | 20.269 | 23.144 | conv1 | 498.268 | 1022.364 | conv1 | 46.025 | 43.506 | conv1 | 43.404 | 43.577 |
| conv2 | 26.661 | 54.883 | conv2 | 304.876 | 659.105 | fire2 | 18.646 | 26.165 | conv2 | 48.861 | 82.239 |
| conv3 | 6.359 | 13.395 | conv3 | 247.751 | 535.662 | fire3 | 18.119 | 26.313 | incep_3a | 34.198 | 53.154 |
| conv4 | 8.420 | 18.624 | conv4 | 132.813 | 281.132 | fire4 | 38.098 | 53.110 | incep_3b | 46.743 | 74.526 |
| conv5 | 8.487 | 19.019 | conv5 | 44.783 | 90.830 | fire5 | 11.464 | 16.732 | incep_4a | 22.026 | 34.657 |
| fc6 | 12.651 | 28.165 | fc6 | 31.676 | 74.651 | fire6 | 14.034 | 20.692 | loss1 | 6.931 | 11.618 |
| fc7 | 6.419 | 13.580 | fc7 | 6.291 | 14.291 | fire7 | 13.851 | 20.989 | incep_4b | 22.307 | 36.201 |
| fc8 | 1.976 | 5.603 | fc8 | 1.906 | 5.601 | fire8 | 20.640 | 30.215 | incep_4c | 23.094 | 36.396 |
| Loss | 1.883 | 0.994 | loss | 1.755 | 0.930 | fire9 | 8.842 | 13.703 | incep_4d | 25.049 | 38.848 |
| | | | | | | conv10 | 8.076 | 10.592 | loss2 | 7.312 | 11.626 |
| | | | | | | loss | 1.448 | 0.785 | incep_4e | 26.757 | 41.106 |
| | | | | | | | | | incep_5a | 15.832 | 23.933 |
| | | | | | | | | | incep_5b | 15.236 | 24.669 |
| | | | | | | | | | loss3 | 2.389 | 3.445 |
| Ave. | 93.230 | 177.527 | | 1270.420 | 2684.860 | | 199.525 | 263.047 | | 341.288 | 516.490 |
| Ave. F->B | 270.79 | | | 3955.400 | | | 462.600 | | | 857.810 | |

The syncedmem class originally defines four status: uninitialized, CPU, GPU and synchronized, highlighted by green. Memory status can be switched by invoking corresponding high-level functions, e.g. *to_cpu/gpu*, and perform data copy as required. Here we create a new status for FPGA, highlighted in blue, and it means the data is at the FPGA DDR memory at current moment. This status can be added into the original pattern by using extended runtime functions, e.g. *to_fpga/cpu*, resulting in a larger topography. Under such a flexible memory management for heterogeneous platforms, we can achieve a function-level or fine-grained synchronization on various platforms with ease and safety within FeCaffe. The size of memory allocation can be calculated firstly, and then the data can be assigned to any device with the memory size required and management flow mentioned above, given any specific functions or operations. Therefore, the proposed architecture is able to support flexible workload partition on various platforms in theory, taking the bottom part of Figure 3 as the example. A number of functions can be performed straightforward on any specific devices respectively. In the meanwhile, it is also able to partition the workload on GPU, CPU and FPGA respectively with memory synchronization. For simplicity, we only test the combination of CPU and FPGA device, e.g. the fallback mechanism on CPU. Note that OpenCL standard utilizes the master-slave model to synchronize between host and devices, but cannot support synchronization between devices directly. Therefore, there has to be twice synchronization if the FPGA data would like to communicate with GPU.

## 4 Result and Analysis

### 4.1 Performance Benchmark

The result is summarized in Table 1. In this study, we take some classical CNN networks, e.g. AlexNet, VGG-16, SqueezeNet and GoogLeNet, as examples, and perform benchmarks including forward, backward, and for-backward on Intel Stratix 10 FPGA development board with Intel OpenCL SDK version of 19.2. The hardware and software of host platform is Intel Core i7-7700K with 4 Cores and 8 Threads and Ubuntu 16.04 respectively. In order for the accuracy during the training, all of data type is FP32 and native floating-point DSP is implemented for the multiplication and addition operations on FPGA. It is important to note that all of processing kernels for network topologies during training and inference are implemented on FPGA in this test. For the Table 1, the convolution also involves a couple of operations associated, e.g. pooling, Rectified Linear Unit (ReLU), Local Response Normalization (LRN), followed by the convolution layers. The *fire* layer consists of squeeze, expand, ReLU and concat operations accordingly defined in SqueezeNet and *inception* layer contains convolution layers of 1x1, 3x3 and 5x5, ReLU, pooling and concat operations in GoogLeNet. Due to limited space, we use convolution, fire and inception to represent those layers. With respect to the time measurement, we utilize Caffe native *time* function to measure the iterations of 100 times with batch size of 1. The forward and backward flow is the normal approach defined by the native Caffe, computing the result from beginning to the last layer, and gradient from the last to the beginning layer respectively. In this work, we use the *train_val* as the model for each network during the performance measurement so that all of layers are required to perform backward calculations, demonstrating great FPGA adaptation but longer process path and time, compared to the *deploy* model. The performance results are listed in terms of forward, backward and for/backward of each network respectively.

### 4.2 Kernel Breakdowns

In order for the further analysis of FPGA and host behaviors during network forward and backward process within FeCaffe, we choose the deepest network, GoogLeNet, and employ profiling tools to provide workload breakdowns in details, as listed in Table 2. Table 2 elaborates all of execution details, e.g. kernels required, and total instance times for each kernel, including memory write and read. The *Efficiency* column has three meanings in Table 2: one is for FPGA DDR bandwidth efficiency during kernel execution, which is an average ratio and has been dynamically measured by FPGA profiling tool; one is for PCIe data transfer efficiency during memory movement between host and FPGA, which is an average ratio and has been dynamically measured by Intel Vtune Amplifier; the last is the ratio of total kernel execution to total for/backward process time. Given the process with batch size of 1, there are 25 kernels used and 960 times of kernel invocations in total, including 198 times for writing data buffer and 3 times for reading data buffer from FPGA to host. The *gemm* kernel is the most frequent operation with 186 times of invocation. Its total kernel execution time is 58.407ms with 77% FPGA DDR efficiency, and thus the average execution time is about 0.31ms for each invocation. Similarly, the *gemv* kernel is used by 69 times with 7.067ms for total execution time and 81% DDR efficiency, resulting in average 0.1ms for each time. Kernels of gemm and gemv have extra optimization with local memory buffer and Single Instruction and Multiple Data (SIMD) directives for vectorization. Using local memory buffer can dramatically decrease the times of DDR memory access required. Here we use maximum DDR bandwidth of S10 board, i.e. 14928MB/s with FPGA logic running at 300MHz, as the reference, to compare the DDR efficiency for each kernel. With respect to the data transfer, i.e. write buffer from host to FPGA and read buffer from FPGA to host, many writing buffer events are trigged by loading convolution and bias weights to FPGA to perform convolution layer by layer. The average PCIe data transfer speed is measured at 1.906GB/s, resulting in efficiency of 10% by taking PCIe Gen3 x16 lanes as the reference, i.e. 15.75GB/s. Finally, all of kernel and data transfer time are summed up, achieving 598.453ms and accounting for 70% of total average forward and backward time measured in Table 1, which implies that there are some software runtime overhead for current CPU and FPGA collaborations, leaving the room to further optimize in the future.

**Table 2: Kernel Statistics within F->B for GoogLeNet**

| Kernels | Instance Count | Total Time (ms) | Efficiency |
|---|---|---|---|
| Ave_pool_B | 3 | 3.184 | 36% (DDR) |
| Ave_pool_F | 3 | 2.902 | 39% (DDR) |
| Col2im | 19 | 31.197 | 54% (DDR) |
| Concat | 72 | 18.015 | 10% (DDR) |
| Bias | 59 | 20.315 | 12% (DDR) |
| Dropout_B | 3 | 0.113 | 10% (DDR) |
| Dropout_F | 3 | 0.104 | 10% (DDR) |
| Gemm | 186 | 58.407 | 77% (DDR) |
| Gemv | 69 | 7.067 | 81%(DDR) |
| Im2col | 98 | 187.418 | 42% (DDR) |
| LRN_Diff | 2 | 18.390 | 43% (DDR) |
| LRN_Output | 2 | 4.699 | 16% (DDR) |
| LRN_Scale | 2 | 4.645 | 34% (DDR) |
| Max_pool_B | 13 | 66.337 | 62% (DDR) |
| Max_pool_F | 13 | 62.989 | 60% (DDR) |
| ReLU_B | 61 | 20.707 | 17% (DDR) |
| ReLU_F | 61 | 21.313 | 10% (DDR) |
| Softmax | 3 | 0.776 | 0% (DDR) |
| SoftmaxLoss_B | 3 | 0.063 | 0% (DDR) |
| SoftmaxLoss_F | 3 | 0.089 | 0% (DDR) |
| Split | 41 | 22.943 | 11% (DDR) |
| Add | 9 | 5.632 | 17% (DDR) |
| Asum | 3 | 0.124 | 0% (DDR) |
| Axpy | 25 | 12.695 | 20% (DDR) |
| Scale | 3 | 0.070 | 11% (DDR) |
| Write_Buffer | 198 | 28.168 | 12%(PCIe) |
| Read_Buffer | 3 | 0.091 | 0% (PCIe) |
| Total | 960 | 598.453 | 70%(F->B) |

Table 3 shows the detail of configuration file in terms of hardware utilization and FPGA frequency after placement for the measurement mentioned in Table 1 and Table 2. Current configuration only occupies 47% and 31% for total BRAM and DSP resources, given more than half of these resource for further optimization. The gemm and gemv kernels are highlighted cause they both are significant kernels and optimized with higher utilization in terms of BRAMs and DSPs so that convolution and full connection layers can be performed with high efficiency.

**Table 3: Hardware Utilization on S10**

| | ALMs | Regs | M20K | DSPs | Fmax |
|---|---|---|---|---|---|
| Gemm | 107K(12%) | 326K | 2338 (20%) | 1037 (18%) | 253 MHz |
| Gemv | 49K(5%) | 116K | 756 (6%) | 130 (2%) | |
| Total | 616K(66%) | 1415K | 5419 (47%) | 1796 (31%) | |

## 4.3 Training Process on FPGA

Forward and backward are necessary parts, in addition to that, the training process also needs weight update mechanism after forward and backward processing. In Caffe framework, the solver class is used to optimize and update the weights so that the weights can be trained gradually to reach the loss target as we defined during the training iterations. There are three main computation-related phases during the weight update process: *normalization*, *regularization* and *compute_update*. In this study, FeCaffe has considered operations mentioned above with different approaches. Normalization and regularization can be supported via combinations of BLAS-based kernels while the computer_update is enabled by kernel designs directly, e.g. SGD, Adam and other common policies. Therefore, it is clearly seen that the most of computation burden during weight update has been deployed on FPGA, and thus the proposed FeCaffe is able to provide sufficient features to support CNN training for target networks. We use OpenCL native profiling tool and Intel Vtune Amplifier to capture the entire training process, by using GoogLeNet with batch size and iterations of 16 and 10, and results are shown in Figure 5 and Figure 4 respectively.

Figure 5 illustrates all of kernels required and their execution time for each kernel dynamically during the entire training process by performance registers and counters on FPGA. Figure 4 demonstrates the system profiling by VTune with the view zoomed in during training process. The CPU running time is highlighted by green and FPGA behavior is colored with pink. More details of kernel tasks can be checked via the task line, with different colors. It is clearly seen that the CPU and FPGA interactivity during the CNN training process, and CPU usage can be reduced when FPGA is executing kernel task. Host memory bandwidth can also be monitored for each kernel invocation. For the training with FeCaffe, users can reuse the traditional Caffe format, e.g. solver settings, prototxt, and commands to experience the training on FeCaffe. Moreover, snapshot function can also be supported, and thus the proposed FeCaffe can be reviewed as a comprehensive

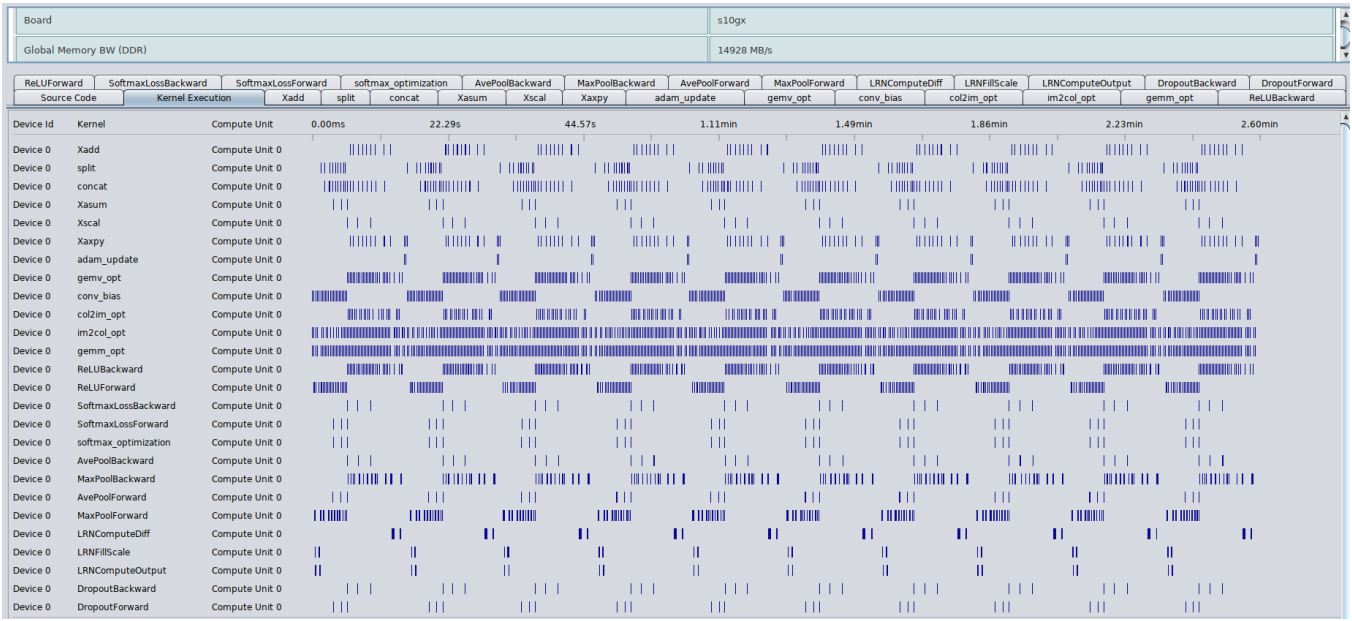

**Figure 5 Kernel Details for GoogLeNet Training Process**

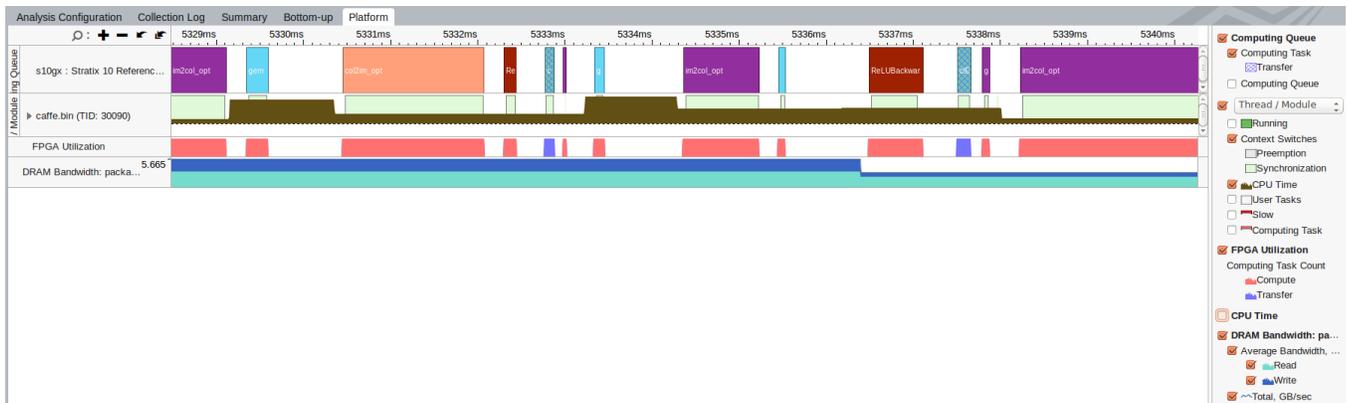

**Figure 4 CPU and FPGA Behaviors duing GoogLeNet Training Process (Best Viewed Zoomed in)**

FPGA-based solution to conveniently provide common deep learning development in particular to training.

### 4.4 Comparison to the State-of-art

This study also presents some comparison with prior work on CNN training with FPGAs in terms of functionality and performance, as listed in Table 4. It is clear that we can provide higher flexibility in terms of CNN network topologies, solver types, training hyperparameter settings, expansibility and ease of use. Compared to FCNN solution [8], we can achieve average execution time of 1102.162 ms and 1710.090 ms for forward and backward respectively, given LeNet with batch size of 384 and 150 minibaches after 200 iterations, resulting in 6.4x and 8.4x average execution time improvement under same testing conditions. Please note that some performance improvement comes from FPGA device difference cause S10 device has native floating-point DSP blocks and more advanced technology node, compared to the device in [8]. Due to DDR memory size limitation of S10 development board, training of VGG-16 and VGG-19 cannot be performed, and thus we provide the training time consumed for one epoch of ImageNet 2012 with 1.2 million training and 50 thousand validation images for AlexNet, SqueezeNet and GoogLeNet, respectively. Compared to FPDeep [9], current training performance is much less competitive. Fitting all of weights, feature data and gradient within on-chip memory over FPGA cluster can significantly change the FPGA pipeline design structure, and maximize FPGA on-chip memory bandwidth and DSP resources, at the cost of 43,200 DSPs and several hundreds of Mbits of BRAMs in total. In addition, fixed-point of 16 is another key factor to provide such an incredible training result. Small batch size is another factor to impact our training speed as total training iterations and data communication times between FPGA and host can be reduced during training and inference phases with the increment of batch size, leading to higher FPGA computation efficiency.

Table 4: Comparison with FPGA Prior Work

| | Our Work | | FCNN [8] | | FPDeep [9] |
|---|---|---|---|---|---|
| Framework | Caffe | | Customized | | Customized |
| Develop Tool | OpenCL with AOC | | MaxCompiler Tool | | RTL Generator |
| CNN Feature | Training and Inference | | Training and Inference | | Training and Inference |
| Network Topologies Supported | AlexNet, VGG, SqueezeNet, GoogLeNet, and the Networks with Same Primitives | | LeNet | | AlexNet, VGG-16 and VGG-19 |
| Solver Supported | SGD, Adam, RMS_Prop, Nesterov, Ada_Grad and Ada_Delta | | SGD Only | | SGD Only |
| Training Hyperparameter Settings | Same with GPUs and CPU, e.g. base_lr, lr_policy, gamma, momentum, weight_decay | | Unknown | | Unknown |
| FPGA Optimization Mechanism | Gemm: NDRange and 2D Local Memory Gemv: NDRange and 1D Local Memory | | Systolic-like: Customized Processing Pipeline for Convolution and Pooling | | All Layers Processing Pipeline Distributed over FPGA Cluster / Store All Weights, Feature and Gradients with on-chip BRAMs / Forward and Backward Processing Pipeline in Parallel |
| Expansibility | Small Efforts to Enable New Functions / No Inter-FPGA Dependency | | More Efforts (Pipeline Need to Update for New Functions) | | More Efforts (Pipeline Need to Update for New Functions) |
| Ease of Use | Same with Conventional Caffe, e.g. Prototxt, Commands and Snapshot | | Customized Network Config. Parameters and HW Constraints | | Customized Network Config. Parameters and HW Constraints |
| Device and Board | Stratix 10 Development Kit | | Stratix V GSD8 | | VC709 Board (V7690T) |
| Number | 1 | | 2 | | 15 |
| DDR Storage and Bandwidth | 2 GB and 14.578GB/s | | 6 GB and 2 * 9.6GB/s | | On-chip Memory Bandwidth |
| Fmax | 253 MHz | | 150 MHz | | Unknown |
| Data Type | FP32 | | FP32 | | Fixed-point 16 |
| Total DSP Utilization | 1796 | | Unknown | | 15 * 2880 = 43,200 |
| LeNet (L1-L6) | Forward (ms) | Backward (ms) | Forward (ms) | Backward (ms) | |
| L1 (Conv) | 524.293 | 514.197 | 590 | 1210 | |
| L2 (Pool) | 22.330 | 23.895 | 530 | 570 | |
| L3 (Conv) | 547.651 | 1156.870 | 4670 | 10320 | N/A |
| L4 (Pool) | 6.539 | 7.010 | 170 | 180 | |
| L5 (FC) | 1.345 | 6.003 | 920 | 1820 | |
| L6 (FC) | 0.004 | 2.115 | 180 | 200 | |
| Total | 1102.162 (6.4x) | 1710.090 (8.4x) | 7060 | 14300 | |
| AlexNet per Epoch | 86.41 Hours (BS:32 and Default Solver) | | N/A | | 0.17 Hour |
| SqueezeNet v1.0 per Epoch | 159.62 Hours (BS:16 and Default Solver) | | N/A | | N/A |
| GoogLeNet per Epoch | 291.08 Hours (BS:16, Default Solver with Adam) | | N/A | | N/A |

## 5 Analysis and Optimization

Based on the result analysis and comparison, the proposed FeCaffe utilizes the fine-grained and kernel-wise FPGA implementation to achieve the same granularity with GPU acceleration, and is capable of providing sufficient and flexible offload functions for deep learning development. It is a new path for deep learning and thus the overall performance is less competitive compared to the mature and well-developed GPU solutions. Therefore, a number of optimization directions mainly focusing on the performance improvement, from FPGA kernel, software runtime, CNN architectures etc., are introduced as follows:

### 5.1 FPGA-level

In this work, we currently choose the OpenCL flow with NDRange format to develop necessary kernels for CNN operations. Due to good adaptation of compiler tool, users are able to deploy most of NDRange kernel files on FPGA conveniently with minor or even no modifications. However, this implementation approach can cause performance issue and resource usage overhead especially for large scale and complicated designs. Therefore, it is recommended by compiler vendors to develop single work-item designs to achieve the best performance with resource optimization. Compared to NDRange style, the single work-item style is very similar with the traditional FPGA design flow, and provides more choices and flexibility to design and optimize kernels. Users can develop more flexible and sophisticated pipeline structures and utilize more optimization directives to fully unleash FPGA massive on-chip memory storage and bandwidth for better throughput performance

Another optimization approach is to improve FPGA logic clock frequency. Stratix 10 FPGA chip has the Hyperflex technology, which inserts some registers on routing resources during placement and routing phase, and thus is able to dramatically increase FPGA design timing frequency [26] . Current implementation approach cannot enable Hyperflex optimization cause this feature only allows single work-item design with stringent conditions for 19.2 version. Therefore, rewriting kernels can increase clock frequency significantly as well. Enlarging DDR storage size and bandwidth for the FPGA board can also improve performance. Currently DDR bandwidth is still a limitation, compared to GPU and CPUs, and thus multiple banks of DDR can mitigate this situation. In addition to these factors, lower bitwidth for training and inference is another important factor to consider for the performance optimization, with the development of retraining and quantization approaches. Int8 and even Int4 can significantly improve DSP efficiency, intermediate data storage and DDR bandwidth and lead to several times of overall CNN processing capability, compared to single floating-point. This enables FPGA solutions to become more competitive compared to GPUs and dedicated ASICs in terms of Int8 and Int4 computation capability.

### 5.2 System Pipeline-level

Currently, FeCaffe chooses synchronous interface to manage communication for higher-level function invocation, and that means the CPU launches FPGA kernels in sequence, and does not start to process the next kernel until current kernel has been completely executed. Therefore, data transfer between CPU and FPGA cannot be overlapped, but is viewed as kernel overhead for the performance measurement. In the meanwhile, FPGA cannot continue to operate all the time cause it has to wait during data transfer, resulting in lower acceleration efficiency. We can notice the phenomenon that kernels are executed discontinuously by Intel Vtune Amplifier in Figure 4. An optimization approach to address this issue is to utilize asynchronous mechanism for CPU and FPGA. By using asynchronous interface, host can put several kernel launches into the invocation queue and thus data transfer through PCIe for next kernels can be prefetched in advanced while FPGA is executing on the current kernel, realizing FPGA continuous operations and higher efficiency. Therefore, overhead of data transfer can be overlapped for the frame throughput calculation and FPGA continuous operation maximizes the throughput performance in terms of system pipeline level.

Fallback on CPU is also viewed as an alternative to improve performance with reasonable workload partition from system level. For example, based on the statistics of kernel execution time for GoogLeNet, total kernel time of *im2col* is the longest, achieving 187.418ms, and the sum of im2col and col2im kernels can reach 218.615ms, accounting for 37% of total kernel time. By nature, the purpose of im2col is to reshape the data and change data address, without data content processing, but requires DDR bandwidth significantly. Therefore, it is wiser to deploy such memory-bounded and small functions on CPU in system workload partition, leaving more burden of data computation at FPGA side.

### 5.3 CNN Network-level

With respect to the current FeCaffe architecture, kernel designs are very fine-grained and fragmented and thus there are enormously frequent communications between CPU and FPGA, e.g. 201 times of memory writing and reading, and 759 times of kernel control to achieve one time of forward to backward for GoogLeNet topology. Therefore, how to reduce interaction times between host and FPGA can significantly improve performance for the given network topology, especially for inference phase. Building large pipeline by merging multiple kernels is a natural and effective approach to mitigate this issue for performance improvement. For this topic, many research studies have been done, as discussed in previous section. For example, graph-based architecture, i.e. pipeline structure to support all of CNN layers on FPGA at one time without CPU runtime interactions, and subgraph-based architecture, i.e. smaller pipeline scale to support layers of convolution, bias, pooling and activation, demonstrate significant advantages in terms of throughput performance and efficiency. Based on the larger pipeline structure, loading weights of entire network can be viewed as the offline initialization stage, and thus interactions of loading weights during runtime can be eliminated. In addition, large pipeline can dramatically shorten the processing latency by leveraging FPGA on-chip memory resource to connect each kernel, compared to normal multiple kernel designs which utilize global DDR memory for data connection between kernels. In order to support and leverage such advanced FPGA hardware architectures within FeCaffe, we can continue to follow the proposed hierarchical approach previously, as illustrated in Figure 6. Hardware designs can be imported and grouped into kernel layer, and runtime functions are required accordingly to manage the kernel operations. Finally, those subgraph-based or graph-based functions can be extended as the customized layers within the operation layer class in FeCaffe to ensure that the underlying optimized architectures can be invoked correctly.

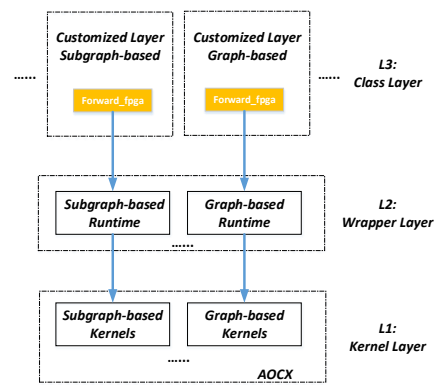

**Figure 6 Integration Approach for Various Architectures**

### 5.4 User Case Application-level

For CNN inference, the throughput performance is a general and straightforward criterion for the quality of various platforms and solutions. Therefore, the ultimate goal of FPGA-enabled CNN inference is to achieve the maximum inference per second with the fixed and optimized pipeline structures. On the contrary, the target of training is to develop new network structures or improve accuracy, forcing algorithm developers to explore and experience novel operations and definitions. Therefore, the flexibility, creativeness and extension capability with ease should be primary factors to consider for the training process. For some scenarios and applications on deep learning-based edge computing and nodes, training is also required from time to time. It is obvious that there is a dilemma by taking into account inference and training designs at the same time on one device. However, for the FPGA, with advanced and systematical reconfiguration mechanism, it has the potential to maintain their own design characteristics of inference and training at the same time.

Partial reconfiguration is an interesting reconfiguration technology for FPGA, and it has high degree of flexibility to allow some functional modification to rapidly update by downloading partial bitstreams while other parts can continue to operate without any interruption [27] . The idea of partial reconfiguration was proposed more than a decade ago, and it has been improving continuously and gradually by FPGA vendors, and becomes mature to work with OpenCL flow in design methodology. Therefore, it can also be considered as one of further optimization direction to the proposed FeCaffe architecture. Based on the discussion above, different targets drive various hardware architectures and thus at least two kinds of architectures, i.e. inference-driven architecture and flexibility-driven architecture for training, or even some variants with tradeoff between performance and flexibility, need to be supported by the system design. In addition, multiple hardware architectures are required to swap on the fly at millisecond level according to use case requirement. Taking into account of these considerations, partial reconfiguration is the most promising choice. The coarse-grained partial reconfiguration-based system for various use case applications is described in Figure 7. Some design modules on memory mechanism can be shared by various reconfigurable designs are placed in static region. Various architectures from inference-driven to flexibility-driven can be complied into a variety of partial bitstreams, which can be managed by host accordingly to various requirements so that FPGA-based deep learning solution can provide efficiency for inference and flexibility for training at the same time.

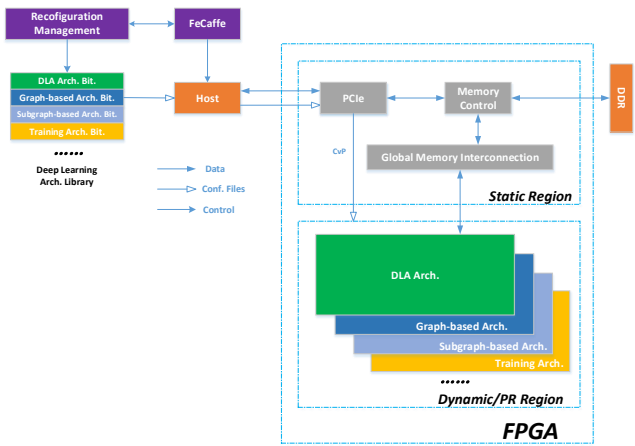

**Figure 7 PR-based FPGA System for User Case Applications**

On the other hand, partial reconfiguration has to bring in more FPGA design complexity and challenges significantly to achieve the benefits discussed above. Firstly, users have to develop and maintain several FPGA architectures for different purposes. Then, they need to plan out some dynamic regions by area constraints on device, and perform FPGA development flow several times to generate various partial bitstreams corresponding to various hardware architectures. Furthermore, how to find out the suitable granularity to fit by taking into account all of kernels from various architectures and build the partial reconfiguration-based pipeline accordingly is the most difficult and challenging for the design. However, it is a positive trend and direction for FPGA designs in CNN applications. Locking to the dedicated and efficient architecture limits FPGA capability, and cannot fully unleash its reconfigurability. This has to force FPGA to become less competitive than deep learning ASIC chips in terms of efficiency, peak performance, power consumption and cost, and thus FPGA-based CNN products have to be displaced automatically and naturally by such emerging and advanced deep learning ASIC chips in the market. However, with the extension of partial reconfiguration-based designs and software management mechanism accordingly, FPGA adaptation and flexibility has been dramatically enhanced. Therefore, FPGA-based CNN products have the potential to support more advanced reconfiguration mechanism to adapt a variety of use cases and scenarios, which is totally and naturally differentiated from current dedicated ASIC chips and conventional GPUs.

### 5.5 Heterogeneous Platform-level

Based on the memory synchronization mechanism between platforms of FeCaffe, creating a hybrid heterogeneous cluster with CPUs, FPGAs and GPUs might be an interesting direction to explore in the future. As discussed in the previous section, the rule of thumb for the cluster is to wisely partition the workload according to various workload characteristics and device features by nature. For instance, latency-sensitive modules are supposed to be deployed on FPGA with single work-item format to build flexible and deep pipeline so that dataflow path can be fully accelerated with FPGA hardware components, achieving the minimum latency to meet the requirements. Memory-hungry or DSP-hungry operations are suitable to offload on GPUs as modern GPUs usually have the highest memory bandwidth in terms of DDR or High Bandwidth Memory (HBM), and have several thousands of processing cores as well. CPU, as the host, is mainly required to manage the synchronization and balance the workload between platforms, and even to perform some small and fragmented functions from time to time, assuming the total execution time of those functions is close to the overhead of kernel launch for GPUs or FPGAs. Therefore, the proposed FeCaffe architecture has the potential to create hybrid platform cluster with more flexibility and higher performance, and to solve some current issues for GPU-only or CPU-only platforms.

## 6 Conclusion

In this paper, we propose FeCaffe framework, an extension of conventional Caffe, with fine-grained and fragmented kernel design on FPGA and OpenCL development flow for deep learning training, and introduce the hierarchical hardware and software design methodology accordingly in details. A number of benchmark results and performance analysis in detail have been provided accordingly. Compared to some prior studies, the proposed architecture demonstrates obvious advantages in supporting CNN network, solver types, training hyperparameter settings, expansibility, flexibility and ease of use for deep learning training development. In addition, current result can achieve 6.4x and 8.4x performance improvement for forward and backward respectively for LeNet. Based on the current performance analysis, we further propose a number of improvement and optimization directions in the future from FPGA-level, system pipeline-level, CNN network-level, use case application-level and heterogeneous platform-level respectively. Taking into account all of these optimizations, the proposed architecture has great potential to provide better system performance, efficiency and higher degree of flexibility for deep learning and CNN development. Therefore, FeCaffe leads to a new horizon of FPGA-based heterogeneous platform for deep learning development by building a bridge between FPGA low-level kernel design and high-level framework directly, and will create more feasibility and choices with gradual optimization and improvement in the future.